\documentclass[sigconf]{acmart}
\AtBeginDocument{%
  }

\setcopyright{acmlicensed}
\copyrightyear{2026}
\acmYear{2026}
\acmDOI{XXXXXXX.XXXXXXX}

\usepackage{booktabs} 
\usepackage{hyperref}

\begin{document}

\title{\lq{}ENERGY STAR\rq{} LLM-Enabled Software Engineering Tools}

\author{Himon Thakur}
\email{hthakur@uccs.edu}
\orcid{0000-0003-4230-9439}
\affiliation{
  \institution{Department of Computer Science \\
  University of Colorado Colorado Springs (UCCS)}
  \country{United States}
}

\author{Armin Moin}
\email{amoin@uccs.edu}
\orcid{0000-0002-8484-7836}
\affiliation{
  \institution{Department of Computer Science \\
  University of Colorado Colorado Springs (UCCS)}
  \country{United States}
}

\renewcommand{\shortauthors}{Thakur and Moin}

\begin{abstract}
The discussion around AI-Engineering, that is, Software Engineering (SE) for AI-enabled Systems, cannot ignore a crucial class of software systems that are increasingly becoming AI-enhanced: Those used to enable or support the SE process, such as Computer-Aided SE (CASE) tools and Integrated Development Environments (IDEs). In this paper, we study the energy efficiency of these systems. As AI becomes seamlessly available in these tools and, in many cases, is active by default, we are entering a new era with significant implications for energy consumption patterns throughout the Software Development Lifecycle (SDLC). We focus on advanced Machine Learning (ML) capabilities provided by Large Language Models (LLMs). Our proposed approach combines Retrieval-Augmented Generation (RAG) with Prompt Engineering Techniques (PETs) to enhance both the quality and energy efficiency of LLM-based code generation. We present a comprehensive framework that measures real-time energy consumption and inference time across diverse model architectures ranging from 125M to 7B parameters, including GPT-2, CodeLlama, Qwen 2.5, and DeepSeek
Coder. These LLMs, chosen for practical reasons, are sufficient to validate the core ideas and provide
a proof of concept for more in-depth future analysis.

\end{abstract}





\maketitle

\section{Introduction} \label{sec:introduction}
A vital class of AI-enabled software systems includes tools supporting the Software Engineering (SE) process, such as Computer-Aided SE (CASE) tools and Integrated Development Environments (IDEs). Modern IDEs, such as the Jupyter Notebook, have already integrated advanced AI capabilities. Consequently, AI has become considerably ubiquitous and accessible throughout the software development process. What has remained largely understudied is the energy-efficiency aspect of the new AI-enabled trends in the modern era of SE. In this paper, we focus on Large Language Models (LLMs) used for code generation, for example, automated suggestions of code snippets, in modern IDEs. We propose a novel approach based on Retrieval-Augmented Generation (RAG) and Prompt Engineering Techniques (PETs).

The contribution of this paper is to answer the following research questions (RQ) through an experimental study: \textbf{RQ1:} Can we reduce LLMs' energy consumption or inference time through Retrieval Augmented Generation (RAG) pipelines? \textbf{RQ2:} How do several selected LLMs with different architectures compare in terms of energy consumption and inference latency? \textbf{RQ3:} Can we find a correlation between the LLM's model size and any possible energy efficiency benefits of using RAG across different LLM architectures? \textbf{RQ4:} Can RAG enable smaller and more resource-efficient LLMs to achieve code generation performance comparable to larger LLMs while maintaining low energy usage?

The remainder of this extended abstract is structured as follows. Section \ref{sec:related-work} briefly reviews the literature. In Section \ref{sec:proposed-approach}, we propose our novel approach. Section \ref{sec:experimental-study} presents a summary of our experimental study. Finally, we conclude and suggest the future work in Section~\ref{sec:conclusion-future-work}.

\section{Related Work} \label{sec:related-work}
Rubei et al. \cite{Rubei+2025} demonstrated that strategic prompt engineering significantly reduces energy consumption in code completion tasks using Llama 3. Their optimal prompt configuration with custom tags reduced energy usage by 99\% for one-shot and 83\% for few-shot prompting compared to baselines, while improving the accuracy of the exact match by up to 44\%. This showed that a thoughtful prompt design could substantially reduce the environmental impact of LLM inference without compromising performance. Moreover, Islam et al. \cite{Islam+2025} showed that LLM-generated code can be up to 450 times less energy-efficient than code written by human developers.

\section{Proposed Approach} \label{sec:proposed-approach}
We propose a novel framework that utilizes RAG pipelines to enhance the energy efficiency and/or inference time for LLM-based code generation or improvement tasks in software development tools or environments. The framework measures the resulting impact on energy consumption of the LLM and the total inference time. The proposed framework integrates several components: i) \textbf{LLM:} Our framework allows selecting from a range of LLMs varying in size and architecture, including GPT-2 (125M parameters), CodeLlama (7B parameters), and Qwen 2.5 (7B parameters). Therefore, we support model portability. The LLM is deployed to support automated code generation and improvements. ii) \textbf{RAG Pipeline:} We implement an embedding-based retrieval mechanism using SentenceTransformers \cite{reimers2019sentence} to identify and incorporate relevant code examples based on natural language queries. Our RAG implementation combines dense vector retrieval with prompt augmentation. For each natural language code description, we perform the following steps: a) \textbf{Embedding Generation:} We encode the query using the SentenceTransformer model (specifically, the all-MiniLM-L6-v2 model) to create dense vector representations. b) \textbf{Similarity Search and Retrieval:} We retrieve the most similar examples from the training corpus based on cosine similarity between embeddings. We use the Facebook AI Similarity Search (FAISS) open-source library for efficient vector search when available, with PyTorch fallback for compatibility across platforms. c) \textbf{Prompt Augmentation:} For RAG-enhanced generation, we augment the input prompt with the retrieved examples. For baseline comparisons, we use only the original query without examples. d) \textbf{Example Selection:} We dynamically select the number of examples to include based on the model's context window, ensuring that we do not exceed maximum token limits. For our main experiments, we typically include 2-3 examples. iii) \textbf{External Knowledge Base:} The RAG pipeline requires an external knowledge base for the search and retrieval of relevant contextual information in order to augment the input prompt as explained above. This knowledge base must be compatible with the specific programming languages supported by the SE tool or environment. iv) \textbf{Energy Monitoring:} We utilize the CodeCarbon \cite{schmidt2021codecarbon} library. Note that we do not rely on the CO2 emission estimates, as it is practically impossible to know the exact source of energy.


\section{Experimental Study} \label{sec:experimental-study}
We use the CodeXGLUE dataset \cite{lu2021codexglue}, specifically the CONCODE text-to-code subset, which contains natural language descriptions paired with corresponding Java code. We also use Kaggle's Natural Language to Python Code dataset for a similar task in Python \cite{kaggle-nl-python-code}. Moreover, the source code is available as open-source software at \href{https://github.com/qas-lab/himon-thakur/tree/main/Rag-Enhanced-PETs-for-LLM-Energy-Consumption-Optimization}{https:// github.com/qas-lab/himon-thakur/tree/main/Rag-Enhanced-PETs-for-LLM-Energy-Consumption-Optimization}. 
Our experimental study confirmed the findings reported by Rubei et al. \cite{Rubei+2025} that well-designed prompts can reduce the LLMs' energy consumption. In addition, we observed a possibility of a reduction in the inference time. In the following, we revisit our RQs and answer them based on the achieved experimental results. \textbf{RQ1.} Based on the results, the answer is yes, this can happen, although with mixed results. The findings showed reductions in energy consumption when using RAG for GPT-2 and CodeLlama. However, DeepSeek Coder and Qwen indicated increased energy consumption with RAG. For the inference time, only CodeLlama showed improvements (25\% faster with RAG), while other models experienced increased inference times. CodeLlama achieved the most promising results with both reduced energy consumption and 25\% faster inference time while drastically improving code quality. Other models showed trade-offs: GPT-2 had a slightly better energy efficiency, but it was slower in inference; DeepSeek and Qwen consumed more energy and time, but they could generally produce higher quality code with RAG. \textbf{RQ2.} The energy consumption levels varied significantly across the selected models, with GPT-2 being the most efficient, followed by CodeLlama, while DeepSeek Coder and Qwen used approximately 3x more energy. For inference time, Qwen without RAG was the fastest, followed by GPT-2, DeepSeek Coder, and CodeLlama. \textbf{RQ3.} The answer is no based on our evidence here. The data showed no clear relationship between the model size and achieving any RAG-based energy efficiency benefits. Of the models studied, only GPT-2 (i.e., the smallest in size among the choices) and CodeLlama showed energy reduction with RAG, while the other models saw increased energy consumption regardless of their size. \textbf{RQ4.} Based on our outcomes in this study, the answer is yes. With RAG, GPT-2 on the Kaggle dataset achieved a code quality score of 0.6, which matched DeepSeek Coder's performance while using approximately 3.5x less energy. This demonstrated that RAG could help smaller, more efficient models achieve competitive code generation quality.

\section{Conclusion and Future Work} \label{sec:conclusion-future-work}
In this paper, we have examined how RAG affects energy consumption and inference speed in code generation across different LLM architectures for SE tools. Our experimental results have shown that the impacts of RAG pipelines varied across the studied LLMs: CodeLlama experienced 25\% faster inference times and substantial quality improvements, while smaller models like GPT-2 showed mixed efficiency results despite modest energy savings. For future work, we intend to use more powerful servers. In particular, cloud servers require some deeper understanding of the cloud provider infrastructure to be aware of any potential unwanted influence on our measurements. In addition, we will incorporate well-established code quality metrics, such as CodeBleu, as well as static and dynamic analyses, inspections, and testing to better assess the code quality. We will also explore a combination of RAG and the Model Context Protocol (MCP). Finally, we plan to explore the energy efficiency implications for quantum computing, such as Python code generation for quantum SDKs.


\begin{acks}
This work is in part supported by a grant from the Colorado OEDIT. In preparing this work, we used generative AI models and tools (e.g., GPT-5), to assist with generating or revising content and code.
\end{acks}


\bibliographystyle{ACM-Reference-Format}
\bibliography{refs.bib}


\end{document}